\documentclass[usegraphicx,useAMS,usenatbib]{mn2e}

\title[Warm intervening gas towards PKS 0548-322]{Tentative detection of warm
intervening gas towards PKS 0548-322 with XMM-Newton\thanks{Based on
  observations obtained with {\it XMM-Newton}, an ESA science mission
  with instruments and contributions directly funded by ESA member
  states and the USA (NASA).}}
\author[X. Barcons et al.]{X. Barcons$^{1}$\thanks{E-mail:
barcons@ifca.unican.es},  F. B. S. Paerels$^{2}$, F.J. Carrera$^1$,
  M.T. Ceballos $^1$, M.Sako$^3$ \\
$^{1}$Instituto F\'\i sica de Cantabria (CSIC-UC), 39005 Santander, Spain\\
$^{2}$Columbia Astrophysics Laboratory, Columbia University, 538W,
  120th Street, New York NY 10027, USA\\
$^{3}$Stanford Linear Accelerator Center, 2575 Sand Hill Road M/S 29,
Menlo Park, CA 94025, USA}
\begin{document}

\date{28 February 2005}

\pagerange{\pageref{firstpage}--\pageref{lastpage}} \pubyear{2005}

\maketitle

\label{firstpage}

\begin{abstract}
We present the results of a long ($\sim 93\, {\rm ksec}$) {\it
XMM-Newton} observation of the bright BL-Lac object PKS 0548-322
($z=0.069$). Our {\it Reflection Grating Spectrometer} (RGS)
spectrum shows a single absorption feature at an observed
wavelength $\lambda=23.33\pm 0.01\, $\AA\, which we interpret as
OVI K$\alpha$ absorption at $z=0.058$, i.e., $\sim 3000\, {\rm
km}\, {\rm s}^{-1}$ from the background object. The observed
equivalent width of the absorption line $\sim 30 {\rm m\AA}$,
coupled with the lack of the corresponding absorption edge in the
EPIC pn data, implies a column density $N_{\rm OVI}\sim 2\times
10^{16}\, {\rm cm}^{-2}$ and turbulence with a Doppler velocity
parameter $b>100\, {\rm km}\, {\rm s}^{-1}$. Within the
limitations of our RGS spectrum, no OVII or OV K$\alpha$
absorption are detected.  Under the assumption of ionisation
equilibrium by both collisions and the extragalactic background,
this is only marginally consistent if the gas temperature is $\sim
2.5\times 10^5\, {\rm K}$, with significantly lower or higher
values being excluded by our limits on OV or OVII. If confirmed,
this would be the first X-ray detection of a large amount of
intervening warm absorbing gas through OVI absorption. The
existence of such a high column density absorber, much stronger
than any previously detected one in OVI, would place stringent
constraints on the large-scale distribution of baryonic gas in the
Universe.
\end{abstract}


\section{Introduction}

Current cosmological models restrict the baryon fraction in the
Universe to a few per cent of its total matter and energy content.
A large fraction of that amount of atoms, ions and electrons is
strongly suspected to reside in the intergalactic medium.
Lyman $\alpha$ clouds (including Damped Lyman $\alpha$ absorption
systems) are seen to dominate the baryon content of the Universe
at high redshift \citep{Storrie00}, but at lower redshifts their
number density and the subsequent contribution to the ordinary
matter content of the Universe decrease.

Detailed simulations of the cosmological evolution of the baryons
invariably show that the Lyman $\alpha$ absorbing gas at
temperatures $\sim 10^4\, {\rm K}$ undergoes shock heating at
lower redshifts and its temperature rises to $10^5-10^7\, {\rm K}$
\citep{Cen99, Cen01,Fang00,Dave01}. The baryons in the warm and
hot intergalactic medium (WHIM) reach 40-60 per cent of the
total baryon budget in the local universe, according to above
simulations.  Indeed, most of these WHIM baryons live in small
sparse concentrations, and only those in the deepest potential
wells (groups and clusters of galaxies) can be seen through their
X-ray emission.  For the remaining of the WHIM, resonance
absorption lines from highly ionized elements (OVI, OVII, OVIII,
NeIX, etc.) are the best chance to detect them
\citep{Hellsten98,Fang00}.

Tenuous gas is best detected by absorption rather than by emission, if
a bright (and featureless) enough source can be found, typically an
AGN. The limiting equivalent width detectable for a resonance
absorption line is ultimately determined by the spectral resolution of
the spectrograph in use (assuming proper channel over-sampling of each
resolution element) but most often limited by the signal to noise
ratio.  Very roughly, for a typical $S/N\sim 10$ spectrum, where each
spectral resolution element is sampled by a few channels, the weakest
absorption line detectable has an equivalent width of the order of the
width of one channel.  For the Reflection Grating Spectrometer - RGS
\citep{denHerder01} that we use in this paper, the spectral resolution
of the first order spectra is around $60\, {\rm m\AA}$, and therefore
with a good quality spectrum we might hope to detect lines as weak as
$10-20\, {\rm m\AA}$. With the slightly higher spectral resolution
grating spectrographs on {\it Chandra}, lines as weak as $5-10\, {\rm
  m\AA}$ can be detected, given sufficient signal to noise.

In the recent years there have been a number of detections of
resonance absorption lines arising in the WHIM, both in the soft X-ray
band (with {\it Chandra} and {\it XMM-Newton}) and in the
Far-Ultraviolet band (with {\it FUSE}).  A local component of the WHIM
was first discovered by \citet{Nicastro02} using {\it Chandra} and
{\it FUSE} data towards PKS 2155-204.  \citet{Rasmussen03} reported on
the discovery of local OVII absorption towards PKS 2155-204, 3C273 and
Mkn 421 with {\it XMM-Newton}. \citet{Fang03} confirmed the
detection of a $z=0$ OVII K$\alpha$ line towards 3C273 with {\it
Chandra}.

The detection of more distant intervening WHIM seen in absorption
is so far very limited. \citet{Fang02} detect an OVIII K$\alpha$
absorber towards PKS 2155-204 at a redshift coincident with an
enhanced galaxy density in a well studied region of the sky.
\citet{McKernan03} detect OVIII K$\alpha$ absorption towards the
radiogalaxy 3C120, which can be interpreted either as intervening
gas at $z=0.0147$ or as gas ejected by 3C120 at a velocity of
$\sim 5500\, {\rm km}\, {\rm s}^{-1}$. The current status of the
detection of features from the non-local WHIM has been recently
compiled by \citet{Nicastro05}. In that work, it is hinted that
WHIM absorption lines can account for the missing baryons in the
low-redshift Universe.

In this paper we present observations of the BL-Lac object PKS
0548-322 at $z=0.069$ \citep{Fosbury76}. Prior to that
\citet{Disney74} had detected the presence of a group of galaxies
at $z=0.04$ around this object, suggesting that PKS 0548-322 might
be a cluster member.  Much more recent work by \citet{Falomo95}
found that there is indeed a cluster (A S549) around this BL-Lac
object, but at its average redshift of $z=0.069$.  X-ray emission
from this cluster is very complicated to detect and measure given
the brightness of PKS 0548-322.

We first use the RGS spectra to search for absorption lines.  The
single unresolved line found is identified as OVI K$\alpha$ at a redshift
intermediate between that of the BL Lac and observer. We then use the
EPIC pn spectrum to help constraining the column density and Doppler
velocity parameter of the absorber.  We conclude by examining the
turbulence of the absorbing gas and its possible temperature under the
assumption of ionisation equilibrium.

\section{The {\it XMM-Newton} RGS data}

\subsection{The X-ray data}

PKS 0548--322 was observed by {\it XMM-Newton} \citep{Jansen01} during
its 520th revolution, starting on the 11th of October of 2002, as part
of the AO-2 science operations (Obsid number is 0142270101).  The
observation lasted for 94,500 sec. According to the analysis of the
light-curves of the various X-ray instruments, the radiation
environment conditions were good and with low background except for
the last $\sim 10$ ks where strong flaring appeared.  The pipeline
products that were delivered to us had been processed with version
5.3.3 of the {\it XMM-Newton} Science Analysis Software (SAS).
However, all data products were reprocessed using version 6.0.0 of the
SAS. In addition, the reprocessing incorporated a number of effective
area corrections with respect to the earlier pipeline processed data.

The {\it RGS1} and {\it RGS2} data were filtered for high background
flare intervals, following the data analysis threads under the {\it
XMM-Newton} SOC pages\footnote{http://xmm.vilspa.esa.es}.  This
resulted in 84,600 s of good exposure time for both
spectrographs. Both spectrographs were operated in High Event Rate
mode.

The spectra of order $-1$ and $-2$ of PKS 0548-322 were detected in
both spectrographs. The 2nd order spectra, however, had very few
counts and were ignored in this work. The resulting 1st order spectra
(in bins of width around $5\, {\rm m\AA}$ in the central part of the
spectra) have a variable signal to noise ratio across the spectrum,
which range from virtually zero below 6\AA\ and above 35\AA\ to a peak
value of around $\sim 4-5$ at wavelengths $\sim 15-20$\AA.

The search for absorption lines in the RGS spectra, needs to start
with the determination of the continuum level.  In order to perform
this, we have smoothed the spectra by grouping both the RGS1 and RGS2
data in bins containing at least 50 counts.  The regions below 7\AA\
and above 35\AA, where the sensitivity of the RGS is very small, have
been excluded from any further consideration.  The determination of a
good local continuum level is crucial to assess the significance of
any putative absorption line, and therefore we have restricted the
continuum fits to patches of the RGS spectra where the residuals do
not show any large trends.

\begin{table}
\caption{\label{tabfits} Fits to various wavelength regions with a
power law absorbed by the Galactic column.}
\begin{tabular}{l c c c c}
\hline
Instrument & $\lambda$ range (\AA) & $\Gamma$ & $\chi^2/d.o.f$ & $N_{knots}^a$\\
\hline
RGS1+2 & $5-35$ & $1.74\pm 0.02$ & $2160.26/1351$ & - \\
RGS1 & $5-35$ & $1.74\pm 0.02$ & $1051.40/648$ & - \\
RGS1 & $5-15$ & $2.00\pm 0.12$ & $196.55/131$ & 10\\
RGS1 & $15-35$ & $1.34\pm 0.05$ & $538.07/458$ & 20 \\
RGS2 & $5-35$ & $1.74\pm 0.02$ & $1059.15/699$ & -\\
RGS2 & $5-20$ & $1.98\pm 0.04$ & $599.63/499$ & 14\\
RGS2 & $24-35$ & $0.98\pm 0.18$ & $239.10/195$ & 14 \\
\hline
\end{tabular}
$^a$ Number of knots used in the spline fit
\end{table}

To this goal, we have fitted separately the RGS spectra at both
sides of the non-functioning CCDs.  The RGS1 continuum has been
determined independently in the 7-12\AA\ and 15-35\AA\ regions,
and the RGS2 continuum has been determined independently in the
7-20\AA\ and 24-35\AA. As it is customary, we have used a single
power law spectrum absorbed by a Galactic column of $2.21\times
10^{20}\, {\rm cm}^{-2}$, but in reality any smooth function could
have been used. (Note that in similar optical/UV studies, splines
are often used to take into account the possible presence of weak
broad emission line complexes). The standard fitting routine {\tt
xspec}, version 11.2 \citep{Arnaud96} was used throughout. 
The fitted parameters, along with the corresponding $\chi^2$
values, are presented in table \ref{tabfits}. It is clear that
although an absorbed power law does not provide a good enough fit
to the full band, this model delivers a much more acceptable fit
when restricted to either side of the non-functioning CCDs in both
spectrometers.  We will elaborate further on the continuum fitting
in Section 2.3.

\subsection{Searching for absorption lines}

Detecting absorption lines in a noisy spectrum is a difficult task. We
have taken the ratio of the data to the continuum model fitted with
the binned data, but in the original non-grouped format, to search for
absorption lines.  We have chosen the code EQWID, kindly provided to
us by Saskia Besier at the University of New South Wales (Australia).
This code searches for significant ``connected'' regions deviating
from the continuum, knowing the spectral resolution. There are only 2
absorption features in the 4 spectral patches that have been explored,
which are detected with significance above 3$\sigma$.  One of them
(whose centroid occurs at 26.12\AA) is actually narrower (50 m\AA)
than the spectral resolution of the RGS1 (60 m\AA), and is just
significant at 3$\sigma$ level. All this suggests that this is a
non-real feature and in what follows we ignore it.  There is,
therefore, only one absorption line candidate detected by the EQWID
software, whose centroid occurs at $\lambda=23.33\pm 0.014\, {\rm
\AA}$, with a FWHM of $74\pm 19\, {\rm m\AA}$ and a measured equivalent
width of $32.3\pm 7.8\, {\rm m\AA}$ as returned by that code (1 sigma
errors).  The significance attributed to this detection is $\sim
3.7\sigma$. Unfortunately the RGS2 data does not cover that region,
due to the failure of the corresponding CCD (number 4) early in the mission.

\begin{figure}
\includegraphics[height=8cm,angle=270]{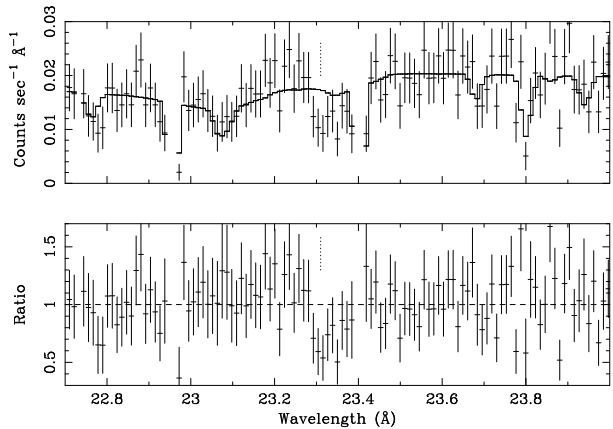}
\caption{A portion of the RGS1 spectrum, fitted to a power law
absorbed by the Galaxy, and the ratio of the data to the model.  The
feature at $\lambda=23.33\, {\rm \AA}$ is the only significant feature
in the whole of the RGS1 and RGS2 spectra that has a FWHM equal or
larger than the spectral resolution.}
\label{spec1}
\end{figure}

Fig.~\ref{spec1} shows a relevant portion of the RGS1 spectrum,
where the absorption line can be seen. There is a feature in the
effective area at $\sim 23.35\, {\rm \AA}$ which \citet{DeVries03}
attribute to an 1s-2p Oxygen transition in the instrument.  This
feature, which occurs at a slightly, but significantly longer
wavelength.  The wavelength discrepancy between the instrumental
absorption feature, and the feature we detect in our source
amounts to $\sim 0.02$ \AA, or slightly more than one third of an
RGS resolution element. The wavelength scale has been calibrated
to much higher accuracy ($8\,$m\AA; den Herder et al 2001). In
addition, as we will show below, the measured equivalent width is
significantly larger than that of the instrumental feature; and
the presence of the instrumental feature is of course explicitly
taken into account in the quantitative spectral analysis. We
conclude that a misidentification of the feature in PKS0548-322 is
excluded.

To further our confidence, we have also inspected other high signal to
noise RGS spectra of extragalactic objects.  \citet{Rasmussen03}
present RGS spectra of 3C273, PKS 2155-304 and Mkn 421. In all 3 cases
an OVII K$\alpha$ absorption line at $z=0$ ($21.60\, {\rm \AA}$) is
seen, with equivalent width $\sim 15-26\, {\rm m\AA}$, i.e., weaker
than our detection.  In these spectra, however, there are no features
detected at around $23.33\, {\rm \AA}$. Section 2.3 presents a further
test, based on the split of our observations in two parts, which
enhances the reliability of the detected feature.

\subsection{More on the continuum and significance of the
absorption line}

All the previous study is based on the assumption that a
power law absorbed by the Galactic column fits well the RGS
spectra on either side of the non-functioning CCDs and that the
calibration (specifically the effective area) is accurate enough.
Indeed, the spectrum of PKS 0548-422 could have slight, local
deviations from the above model and the calibration could still be
affected by small-scale systematics. As it has been mentioned, the
way these issues are solved in optical/UV absorption line studies
is by fitting a local continuum -regardless of its physical
meaning-, often using splines. Visually, the continuum adopted so
far around the putative feature appears a bit low, and this adds
interest to this further analysis.

Therefore we have taken the data to continuum ratio resulting from
the fits to either side of the non-functioning CCDs in both
spectrometers and tried to fit a spline to each one individually.
A region around the absorption line and the instrumental feature
ranging from 23.29 \AA\ to 23.38 \AA\ has been masked out from
this study. In order to prevent over-fitting, that could
eventually follow statistical noise spikes and even potential
absorption or emission features, we explored the decrease of the
$\chi^2$ as a function of the number of knots adopted.  We then
decided to adopt a number of knots below which the $\chi^2$ was
not monotonically decreasing. This number, an obvious function of
the wavelength width of the fitted region, is also shown in table
\ref{tabfits}. In all cases the spline fitting implies a
substantial decrease in the $\chi^2$ fit to a constant. The data
to model ratio for the 4 spectral zones were then divided by the
spline fits and this was then used for further investigation as
explained in what follows.

The routine EQWID was run on the newly re-normalized ratios. Two
features were found with a formal significance above 3$\sigma$.
One at $\lambda=14.70\pm 0.02$ has a too narrow width (FWHM is
only 21 m\AA) to be consistent with a real absorption line and it
is therefore excluded.  The other one, is the same line found
without re-fitting the continuum, with the same formal
significance of 3.7$\sigma$, the same central wavelength
$\lambda=23.33\pm 0.012$, a similar width of $86\pm 19$ m\AA\ and
a slightly larger equivalent width of $39\pm 8$ m\AA. This latter
difference is indeed due to the fact that the continuum fitted by
the splines is slightly larger.  However, we decided to be
conservative and keep the values presented in last subsection,
bearing in mind that the highly uncertain continuum could change
the equivalent width of the line probably upwards.

There is one further experiment we can do with the data to model
ratio re-normalized to the continua spline fits, given the fact
that the distribution of these ratios appears approximately
gaussian. This is to assess the real significance of the line
detected, by means of Monte Carlo simulations.  In order to
perform this, we generated re-normalized data to model ratio
spectra within the same 4 spectral bands, by assuming that values
for each channel are gaussian distributed with mean 1 and
dispersion given by the error bar of the data to model
re-normalized ratio in the real data.  Note that since the signal
to noise ratio varies significantly across the spectra and
therefore we cannot use an overall value for the dispersion of the
gaussian.

We run 10,000 independent simulations, and on each one of them we
run the EQWID software.  We filtered out those features detected
which have either a positive equivalent width (corresponding to an
emission feature) or a width smaller than the spectrometer's
resolution (60 m\AA). With these filters, a total of 605 lines
were detected in the 10,000 simulations with a formal significance
above 3.7$\sigma$.  This in practice means that the probability of
the detected feature at $\lambda=23.33$ being a random noise
feature is 6\%. Indeed this implies a real significance close to
2$\sigma$. We have not, however, inspected every individual
spectrum where a fictitious absorption line has been detected by
EQWID, but a random inspection of a few of them shows that many do
not conform to the shape of an unresolved line and  Voigt profile
fitting either does not converge or results in a very poor fit.
This implies that the real probability that the detected feature
is real is actually higher than 2$\sigma$.

\subsection{Identifying and measuring the absorption line}

\begin{figure}
\includegraphics[height=8cm,angle=270]{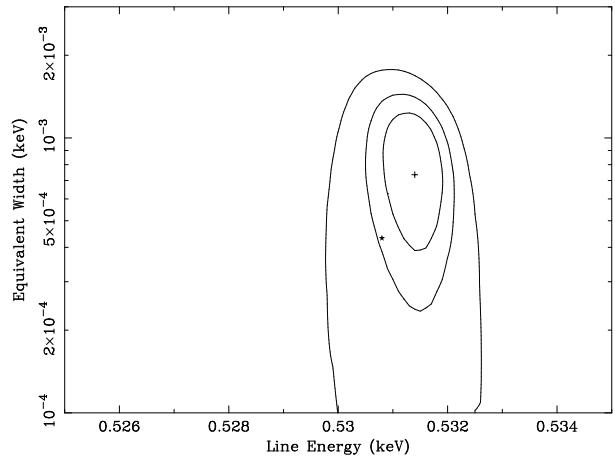}
\caption{Best fit (+ sign) and confidence contours (1, 2 and 3
sigma) for the line energy and equivalent width of a {\it notch}
model fitted to the absorption line. The star sign represents the
instrumental feature reported by \citet{DeVries03}, which is $\sim
2\sigma$ away, and in fact already accounted for in the continuum
model.} \label{notch_params}
\end{figure}

To characterize this absorption feature, we have first modeled the
absorption feature by using the {\it notch} model in {\tt xspec},
added to absorbed power law. This {\it notch} model describes a
fully saturated absorption ``square'' feature, which is good
enough in our case as the line appears unresolved. Inclusion of
this new component in the fit to the RGS1 data in the 15-35 \AA\
range produces a decrease in the $\chi^2_{RGS}$ from 1800.17 (for
1620 degrees of freedom) to 1790.38 (for 1618 degrees of freedom).
The F-test yields a significance of 98.8\% to this new component,
slightly smaller than the significance of the line itself, as
computed by EQWID. We attribute this to the inaccurate modeling of
the line by this simple model.  The best fit is found at a
wavelength of $\lambda=23.33^{+0.028}_{-0.023}\, {\rm \AA}$ and an
equivalent width of $32^{+23}_{-16}\, {\rm m\AA}$ (90 per cent
errors for one single parameter).  Fig.~\ref{notch_params} shows
the best fit and confidence contours for the observed equivalent
width and absorption line energy using the {\it notch} model, both
in keV.

A further test has been performed in order to check whether an
unfiltered bad event (or a set of them) in the region used to subtract
the background could be responsible for the absorption feature. Note,
however, that the shape of the absorption feature conforms to an
approximately unresolved absorption line. To this end we splitted the
RGS1 event file in two approximately equal halves in time, extracted the
corresponding spectra, and computed $\chi^2$ from the previously fitted
models with and without the notch.  The residuals of the model without
the absorption modeled, both show a negative signal around $\sim
23.33\, {\rm \AA}$. Adding the notch model, for the
first half of the observation the $\chi^2$ goes down from 1931.27 to
1926.83 when the absorption "notch" is included with the parameters
fitted to the overall spectrum, and for the second half the $\chi^2$
also decreases from 1895.56 to 1891.64.  Therefore, although each half
of the observation is not sensitive enough for a detection of the
absorption feature, it is clear that a single bad event in the
background region is not producing the observed line.

In order to identify the absorption line, assuming it is caused by
intervening material, we made an extensive search for resonance lines
associated to ground configurations, falling in the wavelength range
from $\sim 23.33\, {\rm \AA}$ (in the case absorbing material occurs
at $z=0$) down to $\sim 21.82\, {\rm \AA}$ (in the case the line
occurs at the redshift of PKS 0548-322 $z=0.069$).  Note that the most
often seen K$\alpha$ transition of OVII (at a rest wavelength of
$21.60\, {\rm \AA}$) falls outside this range by a large amount. In
fact, for the line to be identified as OVII $\lambda 21.60$ the
intervening material would have to be falling towards PKS 0548-322 at
a velocity of $\sim 3000\, {\rm km}\, {\rm s}^{-1}$, which is far too
large for a cluster of galaxies, especially for a poor cluster like
A S549. Note that a significant velocity difference (but in the
opposite sense) has also been found by \citet{McKernan03} in an OVIII
K$\alpha$ absorber towards 3C120. In that case (but not in ours,
if the line is OVII K$\alpha$) there is a possible interpretation
of the absorbing gas as being in the jet and moving towards us.

The only physically meaningful option for that line is that it is OVI
K$\alpha$ $\lambda 22.05$, which has an oscillator strength of
0.576 \citep{Pradhan03}. There is a further line in the doublet at
$\lambda 21.87$, but its oscillator strength (0.061) is far too small to be
detectable. Taking that interpretation, and the central wavelength
measured by the notch model, the redshift of the absorbing gas is at
$z=0.058 \pm 0.001$.

\begin{figure}
\includegraphics[height=8cm,angle=270]{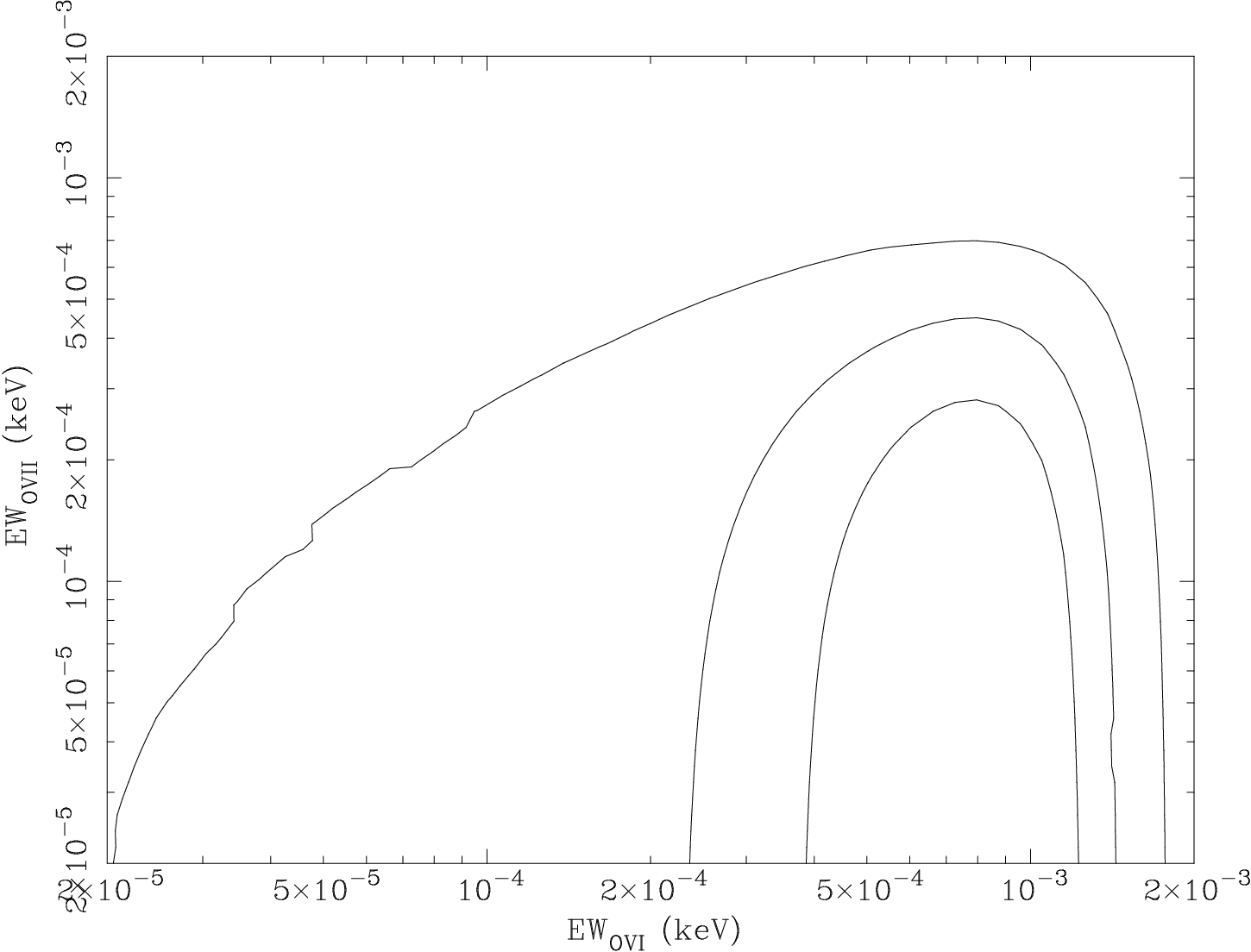}
\includegraphics[height=8cm,angle=270]{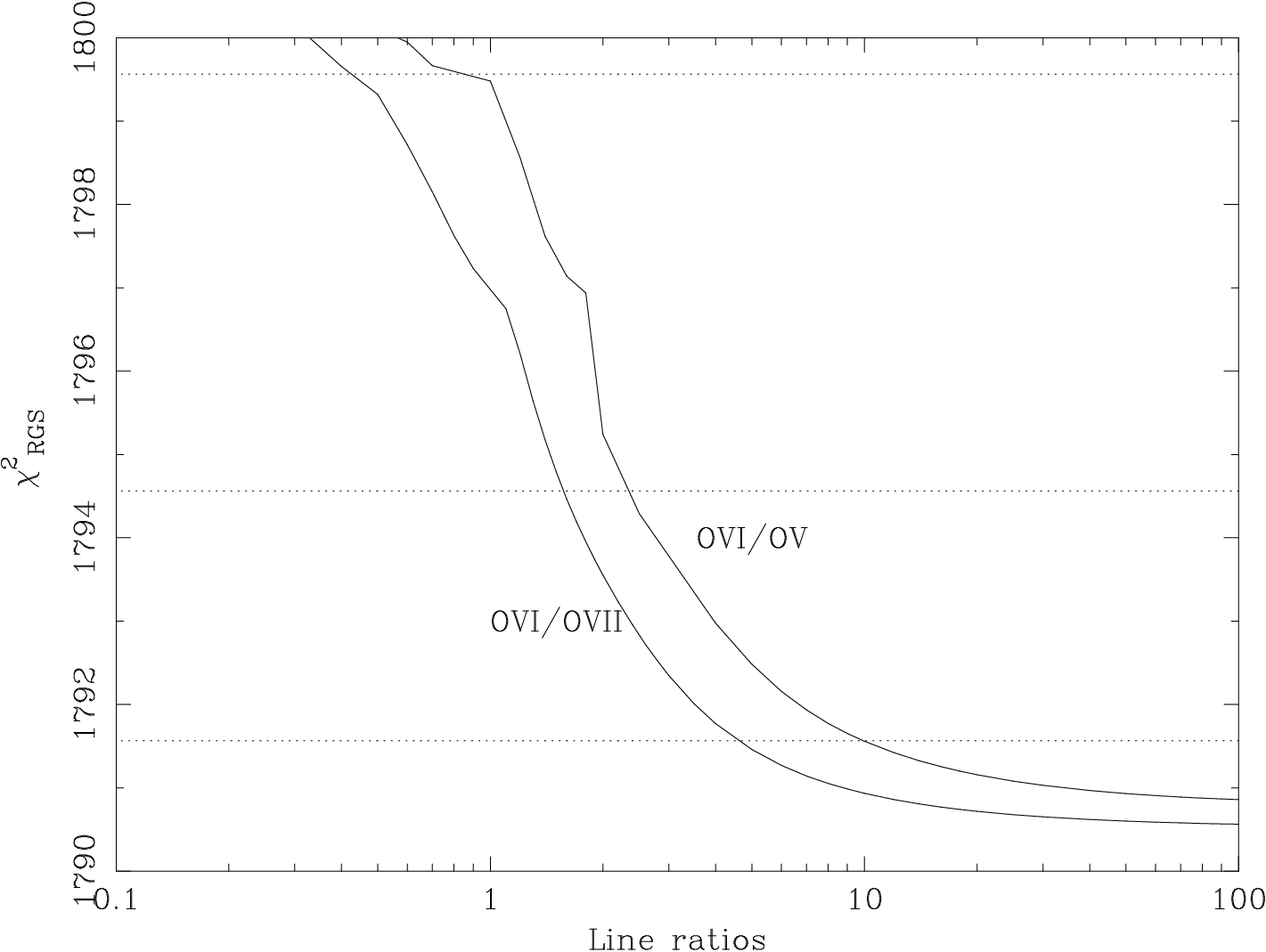}
\caption{Top: Confidence contours in the equivalent width of OVII and
OVI parameter space, as obtained by simultaneously fitting two notches
at the OVI and OVII K$\alpha$ wavelengths and at the same
redshift. Bottom: $\chi^2_{RGS}$ (along with 1, 2 and 3 sigma confidence
levels) for OVI/OVII and OVI/OV equivalent width ratios, as derived from the same
data.}
\label{OVI2OVII}
\end{figure}

Having established the nature of the line at $23.33\, {\rm \AA}$ as
OVI, it is natural to search for the most often seen OVII $\lambda
21.60$ line.  In order to do this, we model the RGS1 data by including
two notches tied down at the same (free) redshift, but assuming that
one is OVI $\lambda 22.05$ and the other one OVII $\lambda 21.60$.
The fit does not improve with the inclusion of the extra notch, but
then we have explored a range in parameter space to establish limits
on the equivalent width of the OVII absorption line.
Fig.~\ref{OVI2OVII} shows the results, first in equivalent width for
OVI and OVII space and then for the OVI/OVII equivalent width ratio.
The conclusion is that OVI/OVII ratio is larger than 4.5, 1.5 and 0.4
at 1, 2 and 3 sigma confidence respectively. We have similarly
constrained the equivalent width of the OV K$\alpha$ line at $22.35\,{\rm
  \AA}$.

\begin{table}
\caption{\label{transitions} Wavelength and oscillator strength of the
  K$\alpha$ transitions for the various Oxygen ions considered here}
\begin{tabular}{l l l}
\hline
Ion & $\lambda$ (\AA) & Oscillator strength\\
\hline
OV & 22.35 & 0.565\\
OVI & 22.05 & 0.576\\
OVII & 21.60 & 0.694\\
\hline
\end{tabular}
\end{table}

\subsection{Profile fitting}

\begin{figure}
\includegraphics[height=8cm,angle=270]{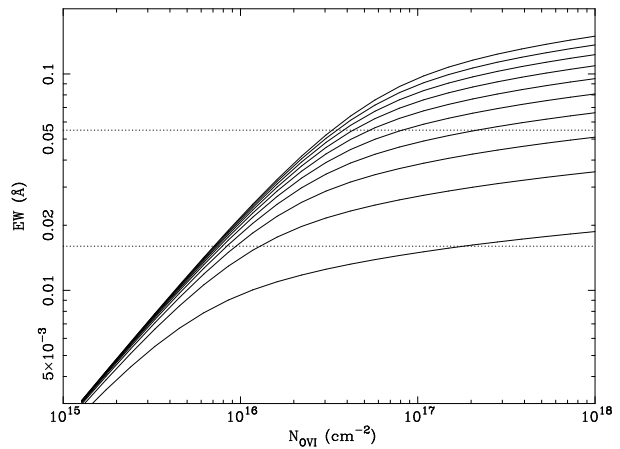}
\caption{Curve of growth for the OVI transition at $22.05\, {\rm
\AA}$. The various curves (from bottom to top) correspond to
Doppler parameters $b$ from 50 to 500 ${\rm km}\, {\rm s}^{-1}$
in $50\, {\rm km}\, {\rm s}^{-1}$ steps.  The horizontal lines
denote the 90 per cent confidence region allowed by the data.}
\label{cogs}
\end{figure}

Fig.~\ref{cogs} shows the curve of growth for the OVI $\lambda 22.05$
transition, for a range of Doppler velocity broadening parameters $b$
($b=\sqrt{2}\sigma_v$, $\sigma_v$ is the velocity dispersion of the
gas, assumed Maxwellian). Although the range of equivalent widths
allowed at 90 per cent level is large, it is clear that some level of
Doppler broadening needs to be present or otherwise the implied value
of $N_{\rm OVI}$ would be very large.

To gain more insight on the physical modeling of the absorption
line, we have performed a Doppler-broadened Voigt profile fitting
to the ratio of the data to model in RGS1.  The Voigt profile has
been approximated with the formula by \citet{Whiting68} (accurate
to better than 5 per cent), the thermal broadening is
parameterized in terms of the $b$ parameter and the spectrometer
response has been modeled as a gaussian of FWHM $750\, {\rm km}\,
{\rm s}^{-1}$ outside {\tt xspec} (i.e., the model spectrum has
been convolved with that gaussian before being compared to the
ratio of data to continuum). Natural broadening (both radiative
and auto-ionising) is only relevant to very high column densities
and it has been ignored here. Fits were restricted to the $22-24\,
{\rm \AA}$ range.

\begin{figure}
\includegraphics[height=8cm,angle=270]{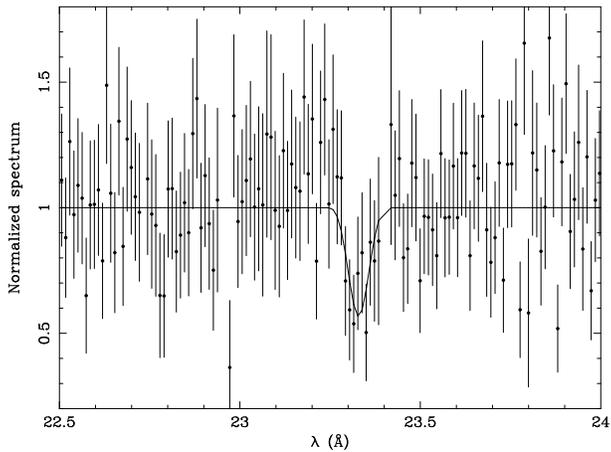}
\caption{Fit to the ratio of the data to the continuum with a
Voigt profile for the OVI line, Doppler broadened with $b=200\,
{\rm km}\, {\rm s}^{-1}$.}
\label{alf-fit}
\end{figure}

Since a fit with the three parameters ($N_{\rm OVI}$, $b$ and $z$)
left free yielded an undefined value for $b$ ($b=435\pm 420\, {\rm
km}\, {\rm s}^{-1}$, 1-sigma error), we fixed $b$ to a reasonable
value of $200\, {\rm km}\, {\rm s}^{-1}$ (see later).  The best fit
provides a substantial improvement of the $\chi^2_{Voigt}$ which goes
from 182.4 for 171 data points to 164.94 with 2 degrees of freedom
less.  The F-test significance of this improvement is at the level of
99.98 per cent.  The best fit values are
$z=0.0580^{+0.0008}_{-0.0006}$ and $\log N_{\rm OVI} ({\rm
cm}^{-2})=16.30 \pm 0.35$ (90 per cent errors). Fig.~\ref{alf-fit}
shows the portion of the spectrum fitted, where it is seen that the
fit is indeed satisfactory.

There is a previous, shorter {\it XMM-Newton} observation of PKS
0548-322 (Obsid number 0111830201), performed on October 3, 2001.  The
RGS spectra (which accumulate $\sim 48$ ks of good exposure time each)
have been presented and analyzed by \citet{Blustin04} which do not
find evidence for any significant absorption or emission features. Our
own analysis of that data is entirely consistent with this. In fact,
including an OVI absorption line with the parameters fixed to the
above values, to the ratio of data to continuum in the 22-24\AA\
region of the shorter exposure spectrum results in an {\it increase}
of the $\chi^2$ from 59.13 (for 59 degrees of freedom) to 66.73.  This
increase is, however, too modest to exclude the possibility that the
line is present in the noisier spectrum.  There is also the
possibility of a transient absorption feature, as discussed by
\citet{Blustin04} for broad absorption features in BL Lacs.

\section{The EPIC data}

The two EPIC MOS detectors \citep{Turner01} operated in {\it Fast
Uncompressed} mode, where the central chip (where the target was) is
read as a single pixel.  Since a proper background subtraction cannot
be performed in this mode, we ignored the MOS data.

The EPIC pn detector \citep{Struder01} was operated in Large
Window mode.  After filtering out high-background episodes, the
net exposure time where data were extracted went down to 77.2 ks.
The spectrum of the source was extracted, following the SAS
analysis threads, using a circle of radius $\sim 40''$.  The
background spectrum was extracted from a $\sim 5$ times larger
region free from obvious sources, although in a different CCD chip
and at a different detector {\tt y} coordinate.

The count rate from the source region is 15.42 ${\rm ct}\, {\rm
s}^{-1}$ (without any pattern filtering), slightly above of the count
rate quoted in the {\it XMM-Newton} Users Handbook of 12 ${\rm ct}\,
{\rm s}^{-1}$ where pile-up starts to be relevant. In order to
quantify the effects of pile-up, we used the SAS task {\it epatplot}
which extracts the spectral distribution of the events with various
patterns (singles, doubles, triples and quadruples) and compares that
to a pile-up free model.  The comparison shows that in the 0.5-2 keV
is very little affected by pile-up, with an overall survival of single
events of $99.2\%$ and an overall survival of double events of
$98.7\%$ in that band.  The differences between the predicted and
observed spectral distributions do not exceed a few per cent at any
particular channel in that band, which is of the order of the accuracy
in the calibration of the EPIC-pn.  We then believe that the EPIC-pn
can be safely used in the 0.5-2 keV band.  Above that energy, and
particularly around 2-3 keV, deviations from the non-piled-up model,
exceed 10-15 per cent and consequently we ignore these data.

\begin{figure}
\includegraphics[height=8cm,angle=270]{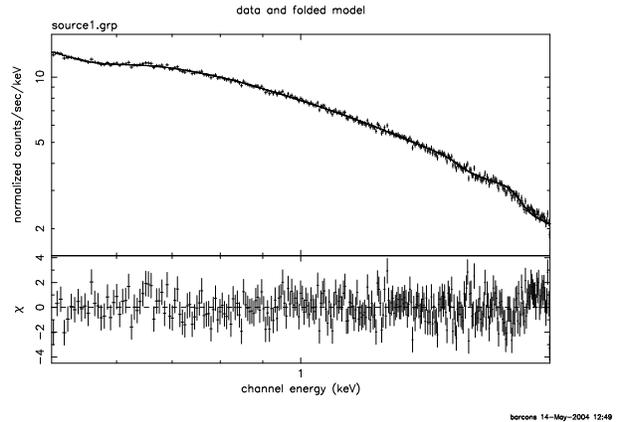}
\caption{EPIC-pn spectrum and deviations to a power law model seen
through Galactic absorption. } \label{specPN}
\end{figure}

Thence we have used the 0.5-2 keV band photons, keeping only single
and double events and only those with good spectral quality ({\tt FLAG
= 0}). A fit to a power law with Galactic absorption fixed at
$N_H=2.21\times 10^{20}\, {\rm cm}^{-2}$ yields a photon spectral
index $\Gamma=2.000\pm 0.005$ (90 per cent errors). The unabsorbed
flux in the 0.5-2 keV band is $S(0.5-2)=(1.865\pm 0.004)\times
10^{-11}\, {\rm erg}\, {\rm cm}^{-2}\, {\rm s}^{-1}$.

Fig.~\ref{specPN} displays the spectral fit and the residuals, where
there are no obvious deviations from the model (none exceeds 5 per
cent at any energy in the $0.5-2$ keV).  The fit is good, with
$\chi^2_{pn}=354.51$ for 302 degrees of freedom, leaving a probability that
the model is wrong of only 2 per cent. Our data does not show the
previously claimed presence of spectral features in this object as
seen by the {\it Einstein Observatory} \citep{Madejski91} and ASCA
\citep{Tashiro94}, which were tentatively interpreted as absorption
edges. Such features, that were once thought to be universal among BL
Lacs, are absent in the current data.

\begin{figure}
\includegraphics[height=8cm,angle=270]{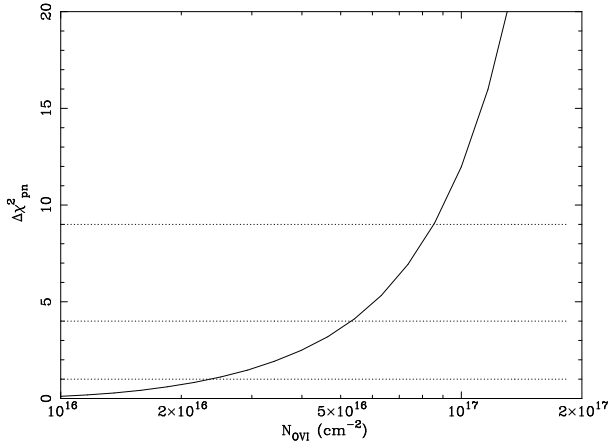}
\caption{$\chi^2_{pn}$ (the minimum has been subtracted) as a function
of the column density $N_{\rm OVI}$ for the fit to the EPIC-pn data in
the 0.5-2 keV band to a power law absorbed by the Galaxy and an OVI
absorber at $z=0.058$, using the {\tt siabs} model \citep{Kinkha03}.}
\label{OVIEPIC}
\end{figure}

We have also tried to search explicitly for the presence of OVI in
this spectrum, by using the {\it siabs} absorption model
\citep{Kinkha03}. This model keeps track of both the photoelectric
absorption edge and the resonance absorption lines (modeled in
terms of proper Voigt profiles) for a given ion species.  Fixing
the redshift of the absorber at the value measured in the RGS data
($z=0.058$) and with a velocity dispersion corresponding to
$b=200\, {\rm km}\, {\rm s}^{-1}$ and with the RGS best fit
$N_{\rm OVI}=10^{16.3}\, {\rm cm}^{-2}$ the $\chi^2_{pn}$ improves
by less than one unit.  Fig.~\ref{OVIEPIC} shows $\chi^2_{pn}$ as
a function of $N_{\rm OVI}$ for the EPIC-pn data, where it can be
seen that the best fit values from the RGS data are within the 1
sigma level (assuming $b\sim 200\, {\rm km}\, {\rm s}^{-1}$).

\section{Properties of the absorbing gas}

\subsection{Turbulence}

Since large values of $N_{\rm OVI}$ are not allowed by the EPIC-pn data,
this poses additional constraints on the curve of growth for OVI. In
particular, a $\sim 30\, {\rm m\AA}$ absorption line resulting from a
large column and very little turbulence (small $b$-parameter) will be
inconsistent with the EPIC-pn spectrum.

\begin{figure}
\includegraphics[height=8cm,angle=270]{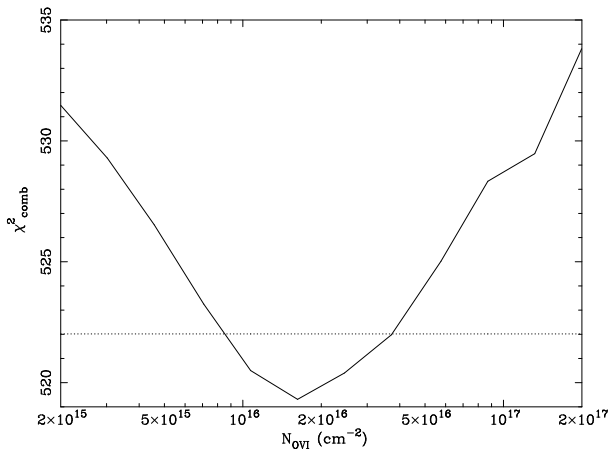}
\caption{$\chi^2_{comb}$ as a function of OVI column density, for a joint fit
  to the EPIC-pn and the RGS1 ratio, where the Doppler velocity
  parameter $b$ has been marginalized.  The dotted line shows the 90
  per cent confidence level.}
\label{chi_N}
\end{figure}

\begin{figure}
\includegraphics[height=8cm,angle=270]{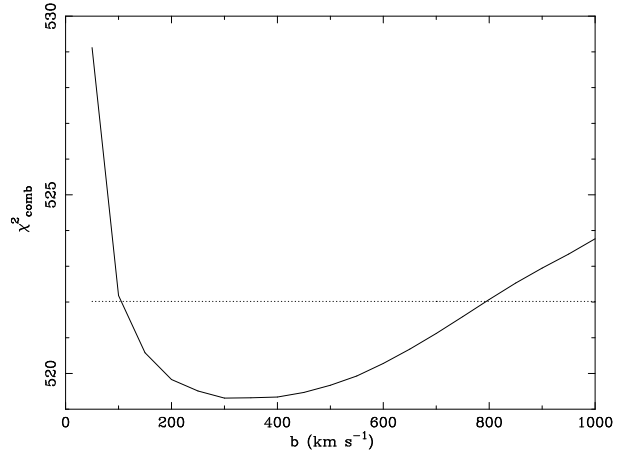}
\caption{$\chi^2_{comb}$ as a function of Doppler velocity parameter $b$, for
  a joint fit to the EPIC-pn and the RGS1 ratio, where the OVI column
  density has been marginalized.  The dotted line shows the 90 per
  cent confidence level.}
\label{chi_b}
\end{figure}

To quantify this, we have explored a grid of values of $N_{\rm
OVI}$ from $2\times 10^{15}$ to $2\times 10^{17}\, {\rm cm}^{-2}$
and for the Doppler parameter $b$ ranging from $50$ to $1000\,
{\rm km}\, {\rm s}^{-1}$.  The RGS1 ratio to the continuum has
been Voigt profile fitted in the range 22-24 \AA\ and the EPIC-pn
spectrum has been fitted in the 0.5-2 keV with a model consisting
of a power law (absorbed by the Galactic H column) and a single
OVI ion as defined by the model {\tt siabs} at $z=0.058$.  For
each value of the parameters $N_{\rm OVI}$ and $b$ the line centre
in the RGS1 ratio is left as a free parameter and the power law in
the EPIC-pn data is also left as free (2 free parameters). We then
minimise
\[
\chi^2_{comb}=\chi^2_{Voigt}+\Delta\chi^2_{pn}
\]

Formally the minimum $\chi^2_{comb}$ is found at $\log N_{\rm
OVI}({\rm cm}^{-2})\sim 16.3$ and $b\sim 300\, {\rm km}\, {\rm
s}^{-1}$. Fig.~\ref{chi_N} shows $\chi^2_{comb}$ as a function of
$N_{\rm OVI}$, coonsidering $b$ an uninteresting parameter, from
which we derive $\log N_{\rm OVI}=16.3\pm 0.3$ (90 per cent
errors). Fig.~\ref{chi_b} shows $\chi^2_{comb}$ as a function of
$b$, where $N_{\rm OVI}$ has been marginalized.  At 90 per cent
confidence level $b=300_{-200}^{+500}\, {\rm km}\, {\rm s}^{-1}$
(i.e., $b>100\, {\rm
  km}\, {\rm s}^{-1}$), but what is most
important, very small values of $b$ are very unlikely.  In fact $b<
50\, {\rm km}\, {\rm s}^{-1}$ is excluded at $>3\sigma$ confidence,
implying that the absorbing gas is turbulent.  Very large values of
$b$ cannot be excluded by this analysis because combined with smaller
values of the column density they can still fit the RGS absorption
line.

\subsection{Temperature}

To test under what physical conditions can the observed limits on
the OVI/OVII and OVI/OV line ratios take place, we performed a
number of {\tt cloudy} runs \citep{Ferland98}. We assume an
optically thin cloud of gas confined by gravity, which is ionised
both collisionally and by photoionisation from the extragalactic
background.  We assumed the gas to be uniform, with 0.3 solar
metallicity, temperature ranging from $10^4$ to $10^7\, {\rm K}$
and particle density ranging from $10^{-6}$ to $10^{-3}\, {\rm
cm}^{-3}$. The particle column density was fixed at some low but
realistic value $10^{20}\, {\rm cm}^{-2}$, but line ratios are
independent on this in the optically thin conditions assumed. We
adopt the simple extragalactic parametrization used by
\citet{Nicastro02}, as a broken power law with energy spectral
indices 1.1 and 0.4 respectively below and above 0.7 keV,
normalized to $10\, {\rm keV}\, {\rm s}^{-1}\, {\rm cm}^{-2}\,
{\rm sr}^{-1}\, {\rm keV}^{-1}$ at 1 keV \citep{Barcons00}.

\begin{figure}
\includegraphics[height=8cm,angle=270]{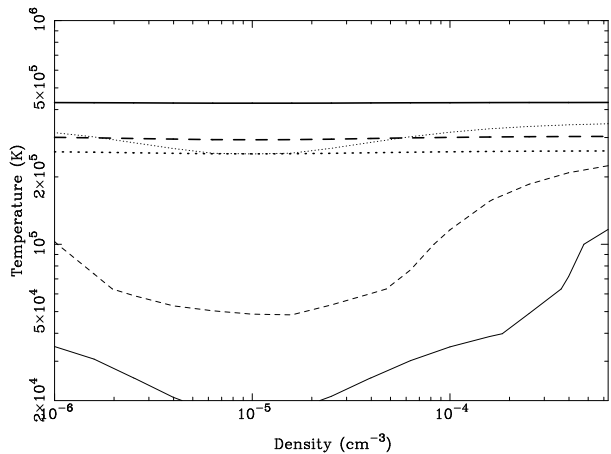}
\caption{Region allowed in the Temperature - Density parameter space
  of the absorbing gas, as obtained from the constraints on the
  OVI/OVII (thin) and OVI/OV (thick) ratios: 1$\sigma$ (solid),
  2$\sigma$ (dashed) and 3$\sigma$ (dotted).}
\label{cloudy}
\end{figure}

Fig.~\ref{cloudy} shows the 1,2 and 3$\sigma$ contours allowed in
the density-temperature parameter space by the measured OVI/OVII
and OVI/OV line ratios.  Note that in displaying these contours,
the line intensity ratios derived from the spectral fitting for
the various K$\alpha$ transitions, have been converted to column
density ratios (which is what {\it cloudy} computes) by dividing
each one by its oscillator strength (see Tab.~\ref{transitions}).
Collisional ionisation dominates at densities $>10^{-5}-10^{-4}\,
{\rm cm}^{-3}$, but at the low density end photo-ionisation by the
extragalactic background is the dominating ionisation process.
The 3$\sigma$ upper limit on the gas temperature for OVI being
seen but OVII being undetected is around $2.5\times 10^5\, {\rm
K}$, fairly independent of the density. This same value of the
temperature is also about the 3$\sigma$ {\it lower} limit for OVI
being detected but OV being undetected.

The conclusion is that it is very difficult for gas under these
conditions to show only OVI K$\alpha$ absorption, without OV and/or
OVII absorption.

\section{Discussion}

The {\it XMM-Newton} RGS1 spectrum of PKS 0548-322 ($z=0.069$)
shows a single absorption feature, formally significant at the
3.7$\sigma$ level. Monte Carlo simulations show that this
significance is probably lower, but in any case higher than
2$\sigma$. Intervening OVI K$\alpha$ absorption at $z=0.058$ is
the most likely interpretation of the absorption line. The
equivalent width is $\sim 30-40\, {\rm
  m\AA}$, a value that is of the order of the one expected to be
produced by a group or cluster. We do not find any significant
absorption feature corresponding to the local WHIM ($z=0$), nor to the
putative cluster surrounding PKS 0548-322 ($z=0.069$).  Interpreting
the detected absorption line in terms of OVII K$\alpha$ absorption at
the cluster redshift, would require the gas inflowing towards the BL
Lac at a velocity close to $3000\, {\rm km}\, {\rm s}^{-1}$, which is
at odds with the expectation that BL Lacs eject material towards the
observer.

Despite previous claims on the presence of strong absorption edges
\citep{Madejski91,Tashiro94}, we do not find such edges in the 0.5-2
keV EPIC-pn spectrum of PKS 0548-322. Adopting the OVI identity for
the absorption feature and combining the spectral fits to the RGS1
spectrum normalized to the continuum (Voigt profile fitting) with the
EPIC pn data, we find that the absorber has an OVI column of $\log
N_{\rm OVI} ({\rm cm}^{-2})=16.3\pm 0.3$, with significant turbulence
(Doppler velocity parameter $b>100\, {\rm km}\, {\rm s}^{-1}$ at 90
per cent confidence).  The lack of OV and OVII K$\alpha$ absorption
lines in the RGS1 spectrum, however, impose almost discrepant
conditions to the temperature of the absorbing gas (assumed in
ionisation equilibrium by collisions and the extragalactic
background). Marginal consistency is only achieved if the gas has a
temperature $T\sim 2.5\times 10^5\, {\rm K}$.

If confirmed this would be the first detection of absorption by
X-ray gas at such a low temperature. Simulations do show that
absorbing gas blobs in the WHIM should span a range of
temperatures between $10^5$ and $10^7\, {\rm K}$ \citep{Dave01},
although so far only higher ionisation species, probably
corresponding to the highest temperatures in that range, have been
detected. \citet{Chen03} have specifically studied the
cosmological distribution of OVI absorbers via simulations. Large
column density systems, such as the one tentatively detected here,
are extremely rare according to the simulations. If confirmed,
this would represent a strong challenge to our knowledge of the
distribution of baryons in Cosmic scales.

The fact that no other features associated either to the intervening
absorber itself or to the local WHIM are not detected, calls for the
need of a better S/N high spectral resolution spectrum to confirm the
existence of this peculiar absorber.

\section*{Acknowledgments}

We are grateful to S. Besier and A. Fern\'andez-Soto for help in
the Voigh profile fitting. Comments from an anonymous referee,
which resulted in substantial improvement of the paper, are also
appreciated. XB, FJC and MTC acknowledge partial financial support
by the Spanish Ministerio de Educaci\'on y Ciencia, under project
ESP2003-00812. FP acknowledges support from NASA, under grant
NAG5-7737.

\bsp

\label{lastpage}

\end{document}